# Oxidation kinetics and non-Marcusian charge transfer in dimensionally confined semiconductors

Ning Xu[1,†], Li Shi[2,8,†], Xudong Pei[3,†], Weiyang Zhang[3], Jian Chen[1], Zheng Han[4], Paolo Samorì[5], Jinlan Wang[2,6,*], Peng Wang[7,*], Yi Shi[1,*], Songlin Li[1,*]

[1] School of Electronic Science and Engineering, National Laboratory of Solid-State Microstructures, and Collaborative Innovation Center of Advanced Microstructures, Nanjing University, Nanjing 210093, China

[2] Key Laboratory of Quantum Materials and Devices of Ministry of Education, Department of Physics, Southeast University, Nanjing 211189, China

[3] College of Engineering and Applied Sciences, Nanjing University, Nanjing 210023, China

[4] Institute of Opto-Electronics, Shanxi University, Taiyuan 030006, China

[5] University of Strasbourg, CNRS, ISIS UMR 7006, 8 allée Gaspard Monge, F-67000 Strasbourg, France

[6] Suzhou Laboratory, Suzhou 215125, China

[7] Department of Physics, University of Warwick, CV4 7AL Coventry, UK

[8] Present address: State Key Laboratory of Organic Electronics and Information Displays & Institute of Advanced Materials (IAM), Nanjing University of Posts and Telecommunications Nanjing 210023, China

[†] These authors contributed equally to this work.

*Corresponding authors. Emails: jlwang@seu.edu.cn, peng.wang.3@warwick.ac.uk, yshi@nju.edu.cn and sli@nju.edu.cn.

## Abstract

Electrochemical reactions represent essential processes in fundamental chemistry that foster a wide range of applications. Although most electrochemical reactions in bulk substances can be well described by the classical Marcus-Gerischer charge transfer theory, the realistic reaction character and mechanism in dimensionally confined systems remain unknown. Here, we report the multiparametric survey on the kinetics of lateral photooxidation in structurally identical $WS_2$ and $MoS_2$ monolayers, where electrochemical oxidation occurs at the atomically thin monolayer edges. The oxidation rate is correlated quantitatively with various crystallographic and environmental parameters, including the density of reactive sites, humidity, temperature, and illumination fluence. In particular, we observe distinctive reaction barriers of 1.4 and 0.9 eV for the two structurally identical semiconductors and uncover an unusual non-Marcusian charge transfer mechanism in these dimensionally confined monolayers due to the limit in reactant supplies. A scenario of band bending is proposed to explain the discrepancy in reaction barriers. These results add important knowledge into the fundamental electrochemical reaction theory in low-dimensional systems.

**Introduction**

The development of modern science and technology benefits vastly from the quantitative understanding on natural phenomena and the acquisition of basic regularities behind the world. In analytical chemistry, however, quantitative understanding on the degradation behavior in bulk substances still represents an attractive but challenging topic. First, it is arduous to fully monitor the degradation processes because they are essentially multiparametric and rely on a plethora of parameters[1–8], such as crystallographic defects, energetic stimuli, and corrosion agents. Some parameters, e.g. the density of atomic defects—a crucial one affecting reaction reactivity, are hard to quantify due to the nature of non-uniform distribution, which would adversely lead to spatial fluctuation of the overall reaction rates among different locations. More formidably, the frontal reaction boundaries naturally evolve into hidden planes inside materials, restricting a direct implementation of reaction tracking. All factors pose grand challenges in quantifying the degradation kinetics in bulk substances.

While two-dimensional (2D) van der Waals materials are being exploited for many emerging applications[9–13], the quantitative information on their environmental stability and service expectancy remains elusive. In contrast to their bulk counterparts, the environmental stability of 2D semiconductors is largely reduced due to the reduction in the vertical dimensionality. In general, they are unusually vulnerable and are subject to degrade on short timescales when exposed in ambient conditions[1,14]. Hence, well formulized oxidation kinetics are not only instructive for practical reliability evaluation, corrosion protection[15,16], and controllable formation of functional layers[17], but also key for understanding the reaction mechanism behind. Structurally speaking, the crystalline 2D monolayer semiconductors represent a unique platform to verify the basic chemical theories that are difficult to fulfill in their bulk counterparts[18,19]. A notable merit is that the density of reactive sites ([V], mainly associated with the number of lattice vacancies) can be readily modulated in 2D monolayers, due to the full exposure of lattice atoms. In addition, the lateral reaction boundaries in 2D systems are simplified as explicit lines exposed outwards and thus are easy to track. Both factors facilitate the quantification of reaction kinetics therein.

To date, most studies on the degradation behavior of 2D monolayers have been carried out via less efficient micro-zone techniques, such as atomic force microscope (AFM)[4–6], transmission electron microscope (TEM)[7,8], and Raman spectrum[2,3,14], which are often spatially limited, time-consuming, or implicit in analysis, to unravel the long-lasting compositional or morphological changes during degradation. In addition, the tunability of environmental parameters (e.g. humidity or temperature) constitutes another requirement for full-parametric quantification. Thus, a convenient strategy that features merits of concurrent spatio-temporal resolutions, adaptive environmental compatibility, and high overall throughputs is highly sought-after.

Here, we report comprehensive survey on the kinetics of photooxidation in 2D $WS_2$ and $MoS_2$ through environment-controlled photoluminescent (PL) imaging inside a glovebox. The ad hoc defect engineering on [V] levels and on-demand controls over the environmental factors allow us to fully access a wide parameter space for quantitative analyses. On the basis of the systematic data, we managed to formulize the lateral oxidation rate as a function of various crystallographic (i.e., [V]) and environmental [i.e., relative humidity (RH), temperature ($T$), and illumination fluence ($F$)] parameters, which thus enables an insight into the roles of lattice vacancies and absorbed aqueous oxygen during degradation. Moreover, we identify the presence of conditional boundaries [V]~$3\times10^{14}$ cm$^{-2}$ and RH~46% that determine the practical oxidation modes (i.e., peripheral or basal) and reaction mechanism (i.e., dry or wet), respectively. Most importantly, through contrastive analyses on the reaction barriers and band alignments in the two structurally identical 2D semiconductors ($WS_2$ and $MoS_2$), we unravel an unusual non-Marcusian charge transfer mechanism in these dimensionally confined systems due to limit in reactant supplies, which represents an important correction to the classical reaction theory. Apart from the technical contributions for reliability evaluation and corrosion protection, these findings also add knowledge into fundamental science.

## Results and discussion

**Methodology and material characterization.** Figure 1a illustrates the schematic experimental setup for the kinetic analysis via real-time PL imaging, where the monolayer (1L) van der Waals semiconductors are globally illuminated with a beam of 455-nm wavelength under an objective lens. Hence, the compositional changes can be conveniently monitored via a real-time CCD camera (See Supplementary Movies 1 and 2). In such a setup, the temporal resolution and spatial breadth are mainly determined by the integration time of the camera and the fields of view of the objective lens, respectively, which were set as 350 ms and 400 μm.

Figure 1b,c shows the serial PL images captured at different reaction stages for two typical oxidation modes (termed as peripheral and basal), in which the local PL quenching occurs due to compositional change after degradation. As a result, the entire reaction process can be well tracked from the progression of the frontal reaction boundaries inward or outward, which are recorded in images captured at planned intervals. We verified that the PL imaging enables extraction of consistent length of reaction ($L_{1L}$) with AFM (Supplementary Fig. 2), constituting a high-throughput way to quantify the kinetic process. The lateral oxidation rate ($r_{1L}$) can be explicitly calculated from the ratio of $L_{1L}$ to illumination time ($t$). An exemplified analysis on the serial PL images given in Fig. 1b is illustrated in Fig. 1d, where $L_{1L}$ exhibits linear dependence on $t$, revealing a roughly constant $r_{1L}$ at $0.16 \pm 0.03$ μm/min.

We note that, in contrast to previous reports on the creation of uniform and flat oxidation products in $WSe_2$ treated with highly reactive ozone[20], the products in this ambient photooxidation process are randomly distributed as amorphous aggregates due to the fluctuation of local strain and temperature. Figure 1e shows the surface morphology for a typical reacted region mixed with pristine and oxidized areas recorded by AFM, where the locally oxidized areas evolve into random protrusions in morphology. X-ray photoelectron spectroscopy (XPS) was used to assess the elemental compositions and valences in the local areas. Before reaction, the pristine areas exhibit intense $2p$ excitations from the sulfur atoms, around 164.0 and 162.8 eV, and $2f$ excitations from the tungsten atoms, around 35.3 and 33.1 eV, corresponding to a valence state of +4 for W atoms in $WS_2$ (Fig. 1f). After reaction, two extra peaks emerge at 36.2 and 38.1 eV, which can be attributed to excitations from the highly valent $W^{+6}$ atoms in derived oxysulfides $WS_xO_y$ or oxides $WO_z$, where the subscripts x, y and z denote the unfixed ratios in the products). The elemental mappings for the typical local areas after peripheral (Fig. 1g) and basal (Fig. 1h) oxidation also support above analyses on the oxidation products, where a noticeable loss (increase) of element sulfur (oxygen) is recorded, in spite of the insignificant change in tungsten content. Hence, the underlying chemical reaction equation during degradation is deduced as

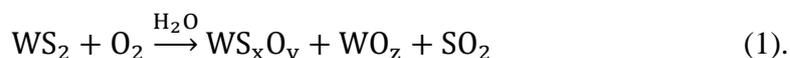

$$WS_2 + O_2 \xrightarrow{\text{H}_2\text{O}} WS_xO_y + WO_z + SO_2 \qquad (1).$$

The material crystallinity in the oxidized areas is completely destroyed and a full amorphous state is detected from the circular STEM diffraction patterns (top inset of Fig. 1i). In contrast, the crystallinity is well preserved in the unoxidized areas, as corroborated by the spotted diffraction patterns (bottom inset of Fig. 1i). Previous theoretical calculations[21,22] revealed that the lattice atoms on the basal planes are more stable than those at edges. Thus, the basal oxidation mode is normally deactivated. Here, by appropriate engineering [V], we reproducibly activated the basal oxidation mode in WS$_2$. Moreover, we identified the threshold [V] level ~3×10$^{14}$ cm$^{-2}$ to activate this mode, as shown in Fig. 1c and Supplementary Fig. 3. Evidently, the full exposure of lattice atoms in 2D semiconductors represents a unique advantage to address the reaction character and mechanism.

**Defect engineering.** A comprehensive survey was performed on the oxidation kinetics versus various parameters, including [V], RH, $T$, $F$ and photon energy (Fig. 2a). In this study, ambiently stable transition metal sulfides, WS$_2$ and MoS$_2$, are selected as the prototypes for their balance in chemical stability and reactivity, because in their pristine states they are robust enough to avoid unintentional degradation before tests, while they can be chemically activated via ad hoc defect engineering[23], such as soaking in oxidants (*e.g.*, solutions of hydrogen peroxide, H$_2$O$_2$, for 5–30 min) to modulate the parameter [V]. In addition, well controlled high [V] levels enable the entire reaction process to accelerate and unfold within a short time of minutes or hours, as compared to, otherwise, months in the pristine states.

Since the precise control and identification of [V] levels are essential for the quantitative characterization, we first prepared defect engineered WS$_2$ monolayers with various [V] levels. Here, we employed an aberration-corrected STEM to estimate the generation rate of [V] versus the pretreatment time ($t_{pt}$) in H$_2$O$_2$ soaking. Figure 2b,c displays the image for a typical 1L WS$_2$ on STEM grid (also Supplementary Fig. 5) and selected atomically resolved image taken under annular dark-field imaging mode. In such a mode, the brightest and adjacent less bright dots correspond to the tungsten and sulfur atoms (denoted by blue and yellow spheres in Fig. 2d), respectively. The lattice vacancies, i.e., loss of atoms (denoted by red arrow in Fig. 2c), would manifest themselves with a further reduced brightness. This feature becomes more evident in the brightness profiling taken along the atom chains (Fig. 2e).

Note that both single- and double-sided sulfur vacancies would correspond to individual reactive sites during the oxidation reaction. The generation rate of reactive sites is then assessed by counting the density of single-sided lattice vacancies per pretreatment time. Figure 2f–j shows typical STEM images for 5 samples, where [V] increases as $t_{pt}$ changes from 10 to 30 min, at steps of 5 min. Atomic images from more local regions and different samples can be found in Supplementary Figs. 8–14. The statistical [V] distributions are all illustrated in Fig. 2k. According to the random nature of vacancy creation on lattice sites, discrete Poisson statistics were used to analyze the spatial distribution of [V]. Theoretically, in Poisson statistics the standard deviation equals the square roots of the average value. We note that the large deviation of experimental [V] levels (~20–30% in most cases) originates from its intrinsic nature, which is inevitable even by increasing sampling numbers, constituting the primary origin of the overall uncertainties in experiment. As can be seen in Fig. 2k, the Poisson statistics well describe the experimental data, implying the reliability of the STEM results. The average [V] level increases linearly from $3.0 \times 10^{13}$ (pristine) to $2.8 \times 10^{14}$ cm$^{-2}$ (30-min-treated). We confirm the structural stability of all the treated samples and exclude any possible phase change from 2H to 1T or 1T' due to the presence of high densities of vacancies, attributable to the random distribution and weak coupling among them, which restrain the collective atomic displacement that is required for triggering a phase change (Supplementary Note 9 and Figs. 16−18). Quantitative XPS analyses on the S/W atomic ratio (Supplementary Fig. 13) also show consistent result of loss of element sulfur after oxidation. Both facilities confirm the efficiency of defect engineering in modulating [V] levels. In Fig. 2l the [V]-$t_{pt}$ trend is plotted, which reveals a generation rate of ~$0.8 \times 10^{13}$ cm$^{-2}$ min$^{-1}$ through soaking WS$_2$ in the 30% H$_2$O$_2$ solution. Overall, this method represents a simple venue to on-demand control [V] levels for quantitative purposes.

**Multiparametric quantification.** The on-demand control over [V] levels allows us to quantify the $r_{1L}$–[V] relationship from experiments. For each condition, we collected relevant $r_{1L}$ data at about 30–120 locations from 3–5 samples (See Supplementary Figs. 20 and 21 for extracting method). Figure 2m shows $r_{1L}$ versus [V] for 1L WS$_2$ at $T = 27°C$ and RH = 60%. At low [V] levels below $1.2 \times 10^{14}$ cm$^{-2}$, $r_{1L}$ is insignificant and approaches the limit of detection (~$10^{-3}$ μm/min), but it increases exponentially when [V] is greater than $1.5 \times 10^{14}$ cm$^{-2}$, indicating the existence of a threshold ~$1.2 \times 10^{14}$ cm$^{-2}$ to trigger fast oxidation of WS$_2$. This result represents fine microscopic evidence for corrosion in 2D disulfides, which well explains the empirical observation between the environmental stability and crystalline quality[24]. A further fitting extracts a characteristic coefficient $\alpha_{1L} = (4.0 \pm 0.5) \times 10^{13}$ cm$^{-2}$ in the exponential correlation $r_{1L} \propto e^{\frac{[V]}{\alpha_{1L}}}$.

Apart from the crystallographic parameter [V], the influences of environmental factors (i.e., RH, $F$, and $T$) on $r_{1L}$ are also quantitatively understood (Fig. 2n–p). Figure 2n shows the dependence of $r_{1L}$ on RH at different $T$s from 20 to 26 ˚C, while Fig. 2o depicts the linear trend of $r_{1L}$ versus $F$ up to 6 W/cm². The $r_{1L}$–RH results are quite informative. An important finding can be attributed to the identification of a general humidity threshold $RH_0 \sim 46\pm4\%$ to trigger appreciable reaction, below which $r_{1L}$ is insignificant at all tested $T$ values from 20 to 26 ˚C. This implies that the 2D disulfides would sustain greatly reduced degradation in surroundings at RH below 46%, constituting useful guidance for corrosion protection. In the broad RH range covered, we also accessed regimes with different response linearity. A linear relation $r_{1L} \propto (RH - RH_0)$ is present in the intermediate RH levels, while $r_{1L}$ becomes saturated at high RH levels. We attribute the deviation of the linear trend at high RH levels to the excessive supply of water molecules, which leads to the saturation of reactants (e.g., the aqueous superoxide anions $O^{2-}$).

To rationalize $r_{1L}$ completely, we further consider the effect of $T$. At first, we exclude the possibility of remarkable $T$ rise induced by photothermal effect. The $T$ rise under typical $F$ levels is simulated to be less than 0.1 K (Supplementary Note 11 and Fig. 22). According to the Arrhenius relation, $r_{1L}$ is plotted versus $1/T$ in Fig. 2p to extract the effective reaction activation energy ($E_a$). Its magnitude is found closely associated with absolute humidity (AH), which divides the inset of Fig. 2p into two regions: I) dry oxidation; II) wet oxidation. In the dry oxidation region, with AH spanning from 9 to 11 g/m³, $E_a$ gradually drops from 2.5 to 1.8 eV, whereas it keeps almost constant at $1.4 \pm 0.3$ eV in the wet oxidation region (AH ranging from 11 to 14 g/m³). This behavior implies the existence of distinct reaction mechanisms and paths that are relevant to environmental humidity.

The extensive data enable us to formulize the oxidation kinetics of 1L WS₂ in a quantitative form. By fitting the data in the linear RH regime, we derived

$$r_{1L} = A \cdot F \cdot \beta \cdot (RH - RH_0) \cdot e^{\frac{[V]}{\alpha_{1L}}} e^{-\frac{E_a}{kT}} \qquad (2)$$

where $A = e^{28\pm1} m^4 \cdot s^2 \cdot (kg)^{-2}$ is a prefactor accounting for parameters not addressed explicitly, $\beta = 2.17$ g·K·J⁻¹·e_s(T)·T⁻¹ is inserted for the algebraic conversion from RH to AH, and e_s(T) is the $T$ dependent saturation vapor pressure, in unit of Pa. For the two exponential terms, the former reflects the crucial role of chalcogen vacancies in the degradation process and the latter is a natural result of the Arrhenius plot to reflect $E_a$. This semi-empirical equation can be used as guidance for evaluating reliability and service expectancy in electronic devices (Supplementary Fig. 23) and for exploiting strategies of corrosion protection in these 2D semiconductors.

**Reaction paths and energy landscapes.** To further shed light on the underlying reaction mechanisms[3,14], we looked into the threshold of photon energy for oxidation reaction. In Fig. 3a, we identify the threshold ~1.94 eV to trigger fast oxidation with appreciable $r_{1L}$, because the oxidation is inactive at energy below 1.91 eV (i.e. above wavelengths of 650 nm, see also Supplementary Fig. 24) and is active when energy equals or exceeds 1.94 eV. We note that this threshold is rather close to the optical gap (~2 eV) of 1L $WS_2$ from the absorption spectrum (upper red line, Fig. 3a), confirming the photochemical reaction mechanism[3,25]. In detail, the role of illumination is to excite electrons from the valence ($E_V$) to conduction band ($E_C$) in $WS_2$; the excited electrons are then transferred to the aqueous oxygen via electrochemical paths to produce superoxide ions $O^{2-}$ as the active oxidant (Fig. 3b,c). In addition to illumination, oxygen and humidity were also found crucial elements to trigger fast degradation. Even illuminated, the samples would behave inactively at low oxygen and humidity levels (Supplementary Fig. 25).

The magnitude of barrier is strongly associated with the crystallographic and environmental conditions, which determine the realistic species of oxidant and reaction path. In Fig. 3b,c we further analyze the possible reaction paths and corresponding reaction energies under four combined cases, in terms of lattice defects (perfect or defective lattices) and humidity (dry or wet conditions), based on the first-principle calculations. Under the dry and wet conditions, the realistic oxidants are molecular $O_2$ and highly active $O^{2-}$, respectively. Combined with the two crystallographic conditions, the reaction barriers are calculated to be 2.89, 2.3, 1.66, and 0.92 eV in the four paths. By comparing the values of barriers, we confirm, at high RH levels, the path of defective lattice plus $O^{2-}$ as the most reasonable one, where the reaction barrier is minimized; the other three paths all feature higher $E_a$ values than experiment. Detailed discussion can be found in Supplementary Note 18.

Apart from the reaction paths, we also investigate the charge transfer process responsible for the photochemical reactions. To this end, the actual energy levels of sulfides were characterized by ultraviolet photoelectron spectroscopy (UPS). The verification of the accuracy of UPS can be found in Supplementary Fig. 27 and 28. Figure 3d shows the UPS spectrum for a p-doped 1L $WS_2$. We extracted a spectral width $W$=15.44 eV, which determines the positions of $E_V$ ~ 5.8 eV and $E_C$ ~ 3.8 eV below the vacuum energy level ($E_{vac}$) after considering the ~ 2 eV energy gap. These derived positions of band edges agree with the theoretical calculations[26].

At the Fermi edge, the UPS spectra feature a characteristic energy of 0.83 eV, giving rise to a Fermi level ($E_F$) at −4.95 eV.

In the classical Marcus-Gerischer theory, an assumption of weak interaction (i.e., a simplified flat band scenario) is often adopted when discussing the charge transfer process in the electrochemical reactions[25,27,28], that is, the energy levels of reactants are assumed virtually isolated and the rule of potential equilibrium between reactants, though being a standard one used in semiconductor physics, is rarely considered. Following this theory, the additional energy barrier is only 0.06 eV for charge transfer from $WS_2$ to absorbed aqueous oxygen, which is too small to interpret the experimental barrier ($E_a$ ~1.4 eV).

Given the positions of energy levels and the magnitude of $E_a$, we propose a strong interaction scenario, that is, the potential equilibrium and band bending should be considered. In Fig. 3e, we plot the actual band alignment during oxidation. Before interaction (Supplementary Fig. 29), the $E_F$ of $WS_2$ lies at −4.95 eV, which is ca. 0.35 eV higher than the redox potential of aqueous oxygen ($E_{F,redox}$ ~ −5.3 eV[25] ). Hence, an upward band bending of 0.35 eV occurs at the reaction interface of $WS_2$, due to the equilibrium of chemical potentials. Accordingly, the photo-excited electrons at $E_C$ need to overcome an extra transfer barrier, $\Delta$, of 0.35 eV to produce the active $O_2^-$ ions. As such, the ion concentration $[O_2^-] \propto e^{-\frac{\Delta}{kT}}$. We further deduced that the two reaction steps are both limited in reactant supplies because of the dimensionally confined nature. Accordingly, the overall reaction rate $r_{1L} \propto [O_2^-] \cdot [TS] \propto e^{-\frac{\Delta}{kT}} \cdot e^{-\frac{0.92\,eV}{kT}}$ , where [TS] denotes the density of transition reaction states for atomic reconfiguration. Hence, the effective $E_a = \Delta + 0.92$ eV = 1.27 eV, which reasonably agree with the experimental value (1.4 ± 0.3 eV). This analysis unravels the crucial roles of potential equilibrium and band bending in understanding the reaction chemistry in dimensionally confined systems.

**Universality of reaction scenario.** To verify the applicability of the above reaction scenario, we also adopted $MoS_2$ as the second prototype. Figure 4a−e shows the corresponding $r_{1L}$ data of $MoS_2$ with respect to the same 5 factors as above, including [V], photon energy, $F$, RH and $T$. By and large, the oxidation characters of $MoS_2$ are quite similar to $WS_2$ (See also Supplementary Fig. 31−38). In particular, its oxidation rate exhibits similar dependence as $WS_2$ for most parameters and can basically share a same analytical formula as Eq. (1), though the relevant coefficients would differ. Surprisingly, both materials share a close humidity boundary at $RH_0$ ~ 46±2% that determines the practical dry or wet reaction mechanism (Fig. 3c).

A few discrepancies are also observed. In general, $r_{1L}$ in MoS$_2$ is much higher than in WS$_2$ (Fig. 4a–e), implying a reduced $E_a$ in MoS$_2$. For this reason, the slope of $r_{1L}$ to RH is too large to access the linear response within our experimental accuracy; only a step-like transition is recorded in $r_{1L}$ as RH crosses the 46% boundary (Fig. 4d). Therefore, the formulized relationship for MoS$_2$ is slightly modified as

$$r_{1L} = A \cdot F \cdot H(RH - RH_0) \cdot e^{\frac{[V]}{\alpha_{1L}}} e^{-\frac{E_a}{kT}} \qquad (3)$$

where H() denotes the Heaviside step function, used to describe its quick saturation behavior. In this case, the prefactor $A = e^{4.2 \pm 0.3}$ Pa$^{-1}$, the characteristic vacancy factor $\alpha_{1L} = 3.7 \times 10^{13}$ cm$^{-2}$, and the activation energy $E_a = 0.9 \pm 0.2$ eV.

It is well known that the WS$_2$ and MoS$_2$ monolayers share several physical features, such as lattice structure and bandgap (~2.0 eV)[26]. Bringing together their $E_a$ data, we then have the chance to further verify the proposed strong interaction scenario, by discussing the origin of $E_a$ difference (1.4 versus 0.9 eV) between them. Following the UPS characterization for WS$_2$, we also determined the energy levels for MoS$_2$ (Fig. 4f) and determined $E_v \sim 6.02$ eV and $E_F \sim 5.20$ eV below $E_{vac}$. We found that there would be hardly solutions to explain the $E_a$ difference between WS$_2$ and MoS$_2$, if the band bending behavior is not considered. For comparison with WS$_2$, we further depict in Fig. 4g the band alignment for MoS$_2$. One can find that the most remarkable difference between the two semiconductors lies in the $E_F$ positions, which are ~−4.95 for WS$_2$ and ~−5.20 eV for MoS$_2$. Because the $E_F$ position of MoS$_2$ (−5.20 eV) is very close to the $E_{F,redox}$ of aqueous oxygen (−5.3 eV), a lower interfacial barrier forms in MoS$_2$ than in WS$_2$ as charges transfer to the surfacial aqueous oxygen. Hence, higher charge transfer capacity and faster oxidation rates are observed in MoS$_2$. This contrastive study emphasizes the role of potential equilibrium in the electrochemical reaction, and adds important corrections to the classical reaction scenario.

In summary, we performed systematic survey on the fundamental photo-oxidation kinetics in 2D transition-metal disulfides, with providing quantitative understanding on the oxidation rate of the monolayer edges versus various crystallographic and environmental parameters. We identified the different conditional boundaries in vacancy level and humidity to trigger different oxidation modes or mechanisms. Also, the microscopic reaction mechanisms were extensively discussed, in terms of the involved reaction paths and energy landscapes. By comparing the actual reaction barriers between WS$_2$ and MoS$_2$, we depicted the realistic charge transfer scenario in the electrochemical process. In particular, we unraveled a non-Marcusian charge-transfer mechanism in the dimensionally confined systems and clarified the crucial roles of potential equilibrium and band bending for quantitatively understanding the reaction barriers, which adds an important correction to the classical Marcus-Gerischer charge transfer theory. Overall, the findings represent profound knowledge about reaction chemistry in low-dimensional materials and also provide instructive guidance for corrosion protection, reliability prediction, and stability evaluation.

## Methods

**Preparation of few-layer disulfides.** Except for the CVD samples used in XPS characterization, all the other $WS_2$ and $MoS_2$ flakes were mechanically exfoliated from synthesized crystals using viscoelastic poly(dimethylsiloxane) (PDMS) as supports. The use of viscoelastic supports helps to achieve large-area and atomically thin flakes with high yields. The ample supplies of 1L disulfide flakes laid the material foundation for this study where a huge quantity of samples were required. To avoid possible contamination, the freshly exfoliated flakes attached on the PDMS supports with clean surfaces (Supplementary Fig. 1) were directly used for defect engineering and oxidation tests without further transfer.

**High-resolution STEM imaging.** The top-view STEM samples for atomic imaging were prepared by transfer of pristine or $H_2O_2$-pretreated monolayers from silicon substrates onto 200-mesh copper STEM grids through polymer poly(methylmethacrylate) (PMMA) as transfer medium (Supplementary Figs. 4 and 5). To enhance the contrast of the sulfur sublattices, a medium-range annular dark-field mode was used. The atomic images were acquired with a double spherical aberration-corrected FEI Titan Cubed G2 60-300 S/TEM. To identify the atomic vacancies while minimizing the adverse bombardment damage (Supplementary Fig. 6) on the fragile monolayers, the STEM was operated under a low acceleration voltage of 60 kV and a probe current of 56 pA. The images used for statistics on vacancy density shown were all processed through the Wiener filtering and average background subtracting. More information can be found in Supplementary Figs. 7 (for $WS_2$) and 32 (for $MoS_2$).

For each condition, 30 independent local regions of an area of $4 \times 4$ $nm^2$ were adopted for [V] statistics (Supplementary Figs. 9–14 for $WS_2$ and Supplementary Figs. 34–37 for $MoS_2$). The independent regions were randomly selected from clean portions free of polymer residues. The spatial distance of the regions ranges from 100 nm to 10 μm, dependent on the surface condition and imaging quality of the samples. On average, 2–5 samples were used for each condition to ensure experimental reproducibility.

**Computational methods.** Density functional theory (DFT) calculations were performed by using the VASP package, to gain insight into the energy landscapes for the possible reaction paths. The electron-ion interactions were evaluated by using the projector-augmented wave (PAW) pseudopotential.[29,30] Exchange-correlation interactions were considered in the generalized gradient approximation (GGA) using the Perdew-Burke-Ernzerhof (PBE) and Heyd-Scuseria-Ernzerhof (HSE). The van der Waals (vdW) interactions were described by using the DFT+D2 scheme. The climbing image nudged elastic band (CI-NEB) method was used to find minimum energy paths. To obtain an ionic pseudopotential for a superoxide anion, we used the configuration $1s^{1.5}2s^22p^{4.5}$, in which a "half" electron from the 1s shell was placed in the valence shell $2p$ of each oxygen atom by an ionic PAW pseudopotential approach. A Monkhorst-Pack k-mesh of $3 \times 3 \times 1$ grid was used to sample the Brillouin zone, and the plane-wave cut-off energy was set to 450 eV. All the structures were fully relaxed until the force on each atom was less than 0.02 eV/Å and the convergence criterion for the electronic structure iteration was set to $10^{-4}$ V. The vacuum space was set to 16 Å in the z-direction, which was large enough to avoid the interactions between two periodic units.

## Data availability

Relevant data supporting the key findings of this study are available within the article and the Supplementary Information file. All raw data generated during the current study are available from the corresponding authors upon request.

## Acknowledgments

This work was supported by the National Key R&D Program of China (Grant Nos. 2022YFA1203802 and 2021YFA1202903, S.L.), the National Natural Science Foundation of China [Grant Nos. 92264202, 61974060, 61674080 (S.L.), 22033002, and 92261112 (J.W.)], the Basic Research Program of Jiangsu Province (Grant No. BK20222007, J.W.), and the Innovation and Entrepreneurship Program of Jiangsu Province. The authors also thank the computational resources from the Big Data Center of Southeast University and the National Supercomputing Center of Tianjin.

## Author contributions

S.L. conceived and supervised the experiments. N.X. performed the degradation experiments and electronic measurements. L.S and J.W. performed the theoretical calculations. X.P, W.Z. and P.W. performed the STEM characterization. J.C. performed the EBL operation. N.X., S.L., Z.H., Y.S., and P.S. analysed the data and discussed the results.

## Competing interests

The authors declare no competing interests.

**Figure caption**

**Fig. 1 | Photoluminescence-traced photo-oxidation process in monolayer (1L) WS$_2$. a**, Schematic setup of real-time photoluminescent imaging. **b,c,** Peripherally and basally initiated oxidation reactions with reaction diagrams depicted on the right. Blue and yellow spheres: W and S atoms, respectively; arrows: directions of reaction progression. The crystallographic and environmental conditions: (**b**) $T$=26 °C, RH=60%, [V]=2.75×10$^{14}$ cm$^{-2}$, $F$=6 W/cm$^2$; (**c**) $T$=22 °C, RH=48 %, [V]=2.75×10$^{14}$ cm$^{-2}$, $F$=6 W/cm$^2$. $T$: temperature; RH: relative humidity; [V]: density of reactive sites; $F$: illumination fluence. **d**, Oxidation length ($L_{1L}$) and rate ($r_{1L}$) versus illumination time ($t$). Error bars: standard deviations. Dashed lines: linear or constant fitting. **e**, Contrastive surface morphologies for typically pristine and oxidized areas in a partially oxidized sample, where the oxidized areas evolve into bulged protrusions. **f**, X-ray photoelectron spectra for pristine and oxidized flakes, where excitation from highly valent W$^{+6}$ emerges after oxidation. **g,h**, Transmission electron microscope (TEM) images and corresponding elemental mappings for two partially oxidized flakes that exhibit (**g**) peripheral and (**h**) basal oxidation modes, respectively. Dashed white lines: reaction boundaries. The golden square in **h** denotes the area used for crystallinity analyses. **i**, Crystallinity analyses on the local pristine (crystalline) and oxidized (amorphous) areas from electron-diffraction patterns. The red and blue dots mark the testing spots.

**Fig. 2 | Multiparametric survey on oxidation kinetics in 1L WS$_2$. a**, Various parameters adopted in experiment, including RH, $F$, [V], $T$, $E_{photon}$. **b**, Z-contrast image for a 1L WS$_2$ on STEM grid in low magnification. Insets: enlarged atomic image and corresponding diffraction pattern for the rectangle area. **c**, Atomically resolved annular dark-field image showing an individual sulfur atomic vacancy (denoted by red arrow). **d**, Schematic atomic structure. **e**, Brightness profile for the marked area in (**d**) to identify an atomic vacancy. The blue and yellow spheres denote the W and S atoms, respectively. **f–j**, Typical atomic images for WS$_2$ flakes pretreated for various pretreatment time ($t_{pt}$) values ranging from 10 to 30 min in steps of 5 min, where [V] increases with increasing $t_{pt}$. **k**, Histograms and Poisson statistics on [V] in samples pretreated under different $t_{pt}$ conditions. For a Poisson probability distribution, the standard deviation (D) roughly equals to the square root of the average value of a variable, which thus features a general broad [V] distribution by nature. **l**, [V]–$t_{pt}$ relation extracted from **k**.. **m–p**, $r_{1L}$ versus various parameters: **m**, [V]; **n**, RH; **o**, $F$; **p**, $T$. Inset of **m**: corresponding linear-linear plot; Horizontal dashed line and gray shaded area in **m**: limit of detection and

regimes below. The yellow and blue shaded areas in **n** denote the subthreshold and linear reaction regimes, respectively. Dashed lines from **m** to **p**: linear fittings for relevant data. Inset of **p**: Fitted activation energy ($E_a$) under different absolute humidity (AH) levels. Standard deviations are used as the error bars in all panels, if any.

**Fig. 3 | Microscopic oxidation mechanisms with simulated reaction paths and derived band alignment. a,** Dependence of $r_{1L}$ on photon energy (left vertical axis) and absorption spectrum of $WS_2$ (right vertical axis). Yellow and blue shaded areas: subthreshold and active regimes of reaction, respectively. Horizontal dashed line and gray shaded area: limit of detection and regimes below. Green and blue symbols: data recorded at AH levels of 16.8 and 19.6 g/m$^3$, respectively. Error bar: Standard deviation. **b,** Two alternative reaction paths for defect-free $WS_2$ with molecular $O_2$ and activated $O_2^-$. **c,** Two alternative reaction paths for defective $WS_2$ containing sulfur vacancies with $O_2$ and $O_2^-$. In **b** and **c**, the blue and red lines represent the energy levels of the initial state (IS), transition state (TS), intermediate state (MS), and final state (FS) during the oxidation reaction. Simulated atomic configurations during different reaction states are shown below. **d,** Ultraviolet photoelectron spectroscopy spectrum for 1L $WS_2$ collected under electrostatic charge compensation mode. The position of valence band can be estimated from the difference of excitation energy ($hv = 21.22$ eV) and spectral width ($W = 15.44$ eV). Inset of **d**: zooming in near the Fermi edge. $E_V$: valence band; $E_{Vac}$: vacuum energy level. **e,** Charge-transfer mechanism within the framework of Marcus-Gerischer theory for oxidation reaction by considering the carrier equilibrium and band bending between 1L $WS_2$ and water/oxygen redox couple. $E_C$: conduction band; $E_F$: Fermi level; $E_{F,redox}$: redox potential of aqueous oxygen; $D_{red}$: density of states of reduced species; $D_{ox}$: density of states of oxidized species.

**Fig. 4 | Oxidation kinetics in 1L $MoS_2$.** $r_{1L}$ as functions of various crystallographic and surrounding parameters: **a,** [V]; **b,** incident photon energy; **c,** F; **d,** RH; **e,** T; Inset of **e**: Fitted $E_a$ under different AH levels. Yellow and blue shaded areas in **a**: the regimes of linear response and saturation, respectively. Yellow and blue shaded areas in **b** and **d**: subthreshold and active regimes of reaction, respectively. Error bar: Standard deviation. **f,** Ultraviolet photoelectron spectroscopy spectrum for 1L $MoS_2$ collected under electrostatic charge compensation mode. Inset: zooming in on the region near Fermi edge. **g,** Derived band alignment to show charge transfer mechanism between $MoS_2$ and absorbed aqueous oxygen where only a small band bending ~0.1 eV occurs at the reaction boundary.

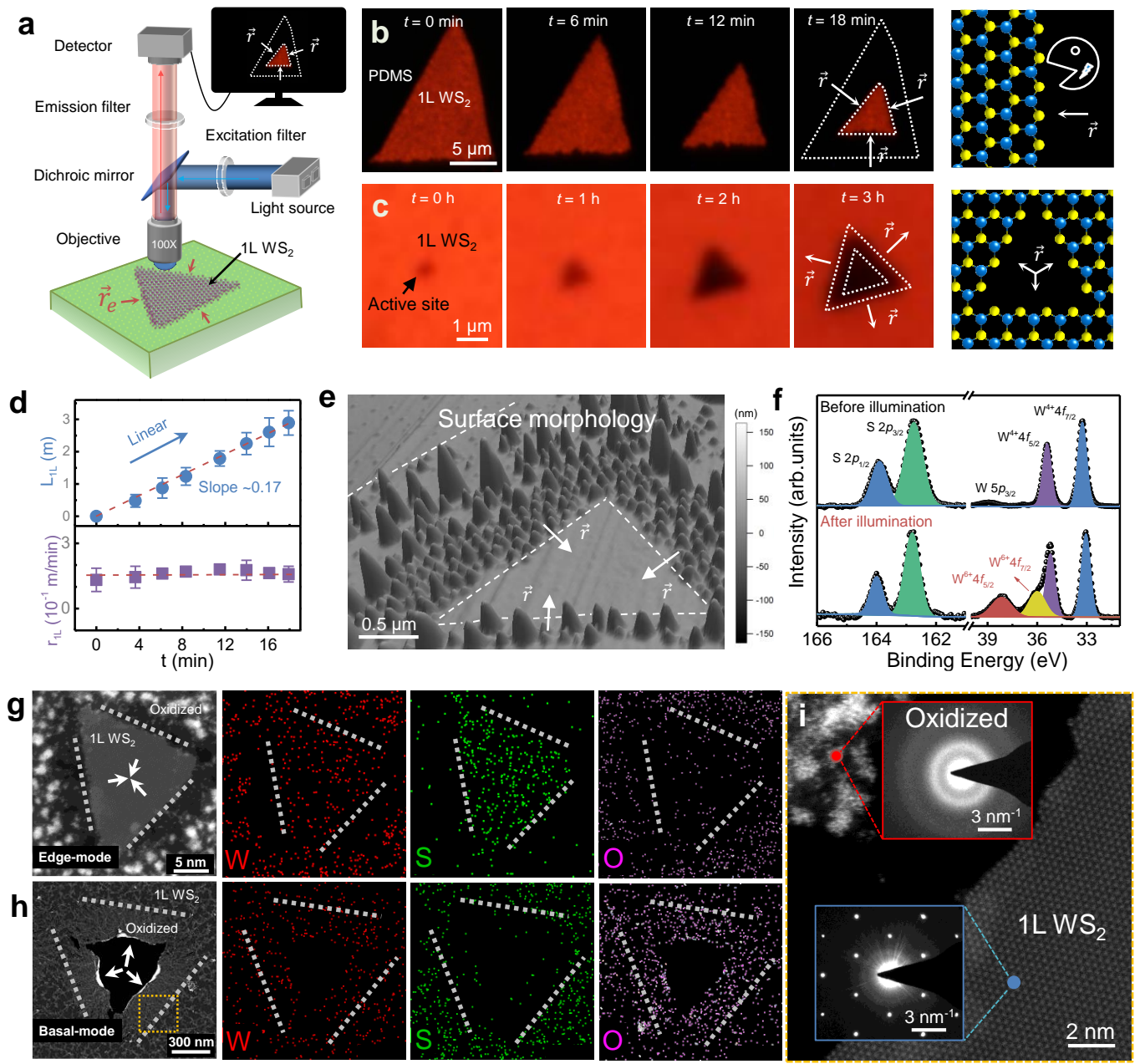

Figure 1

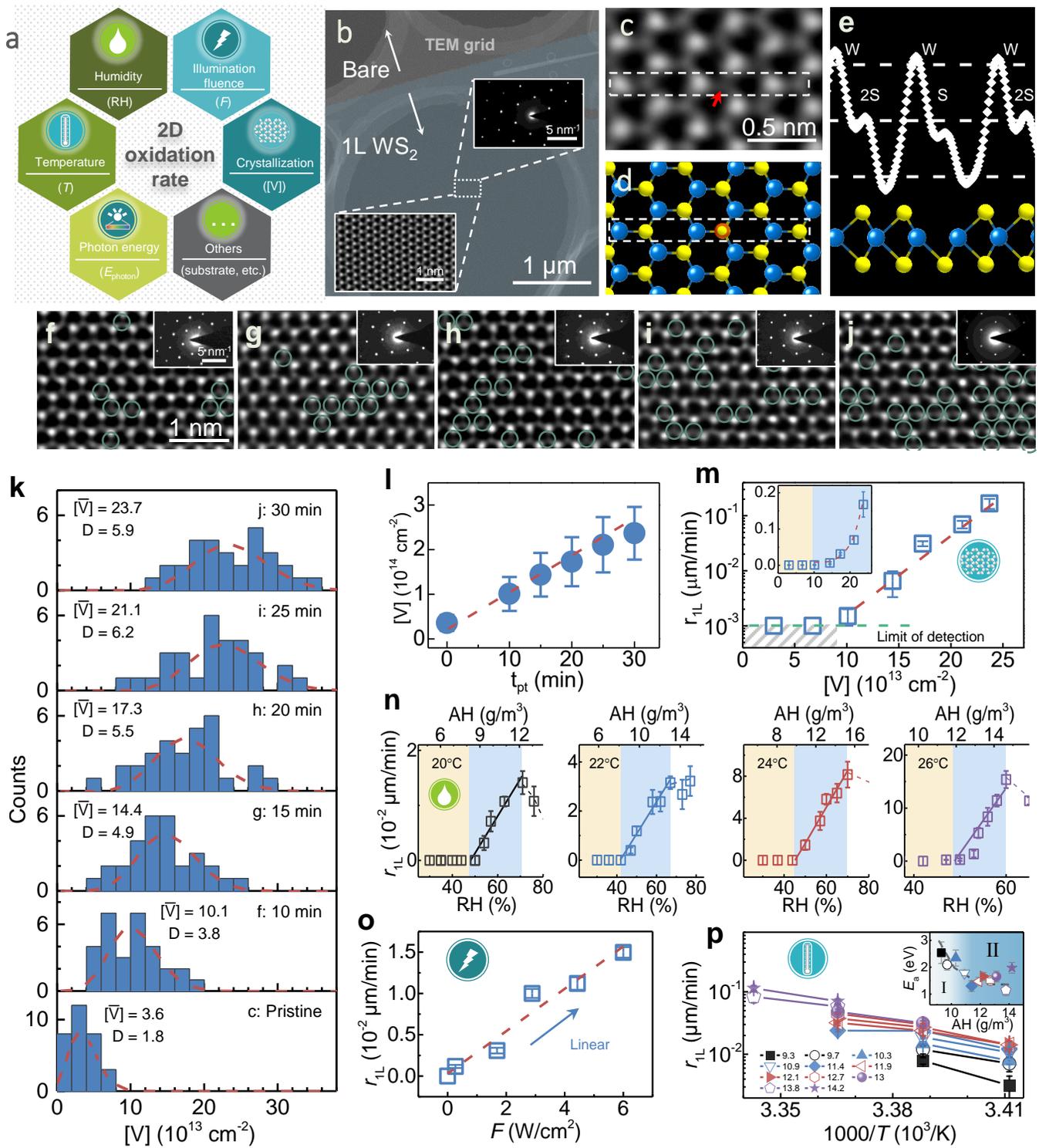

Figure 2

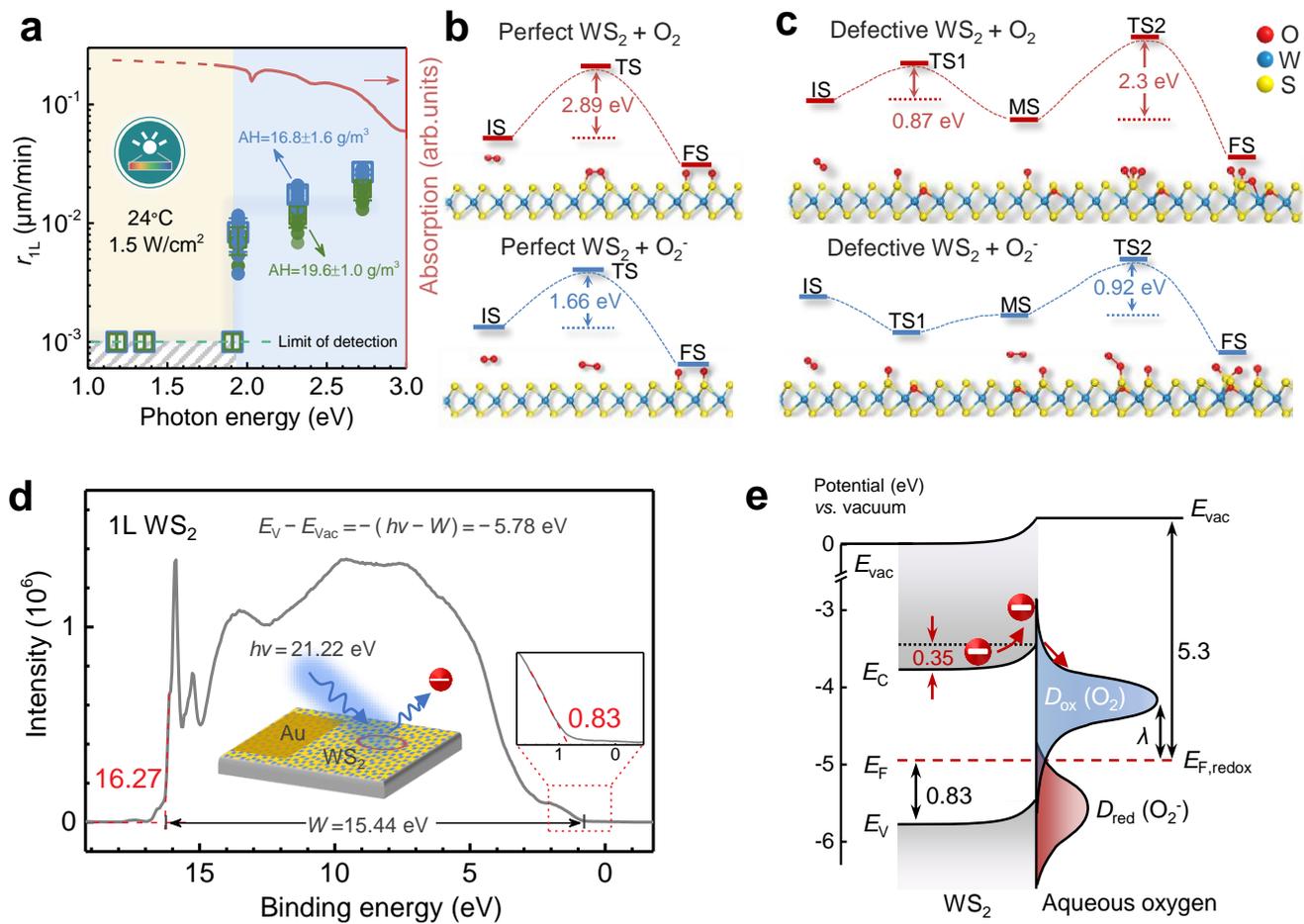

Figure 3

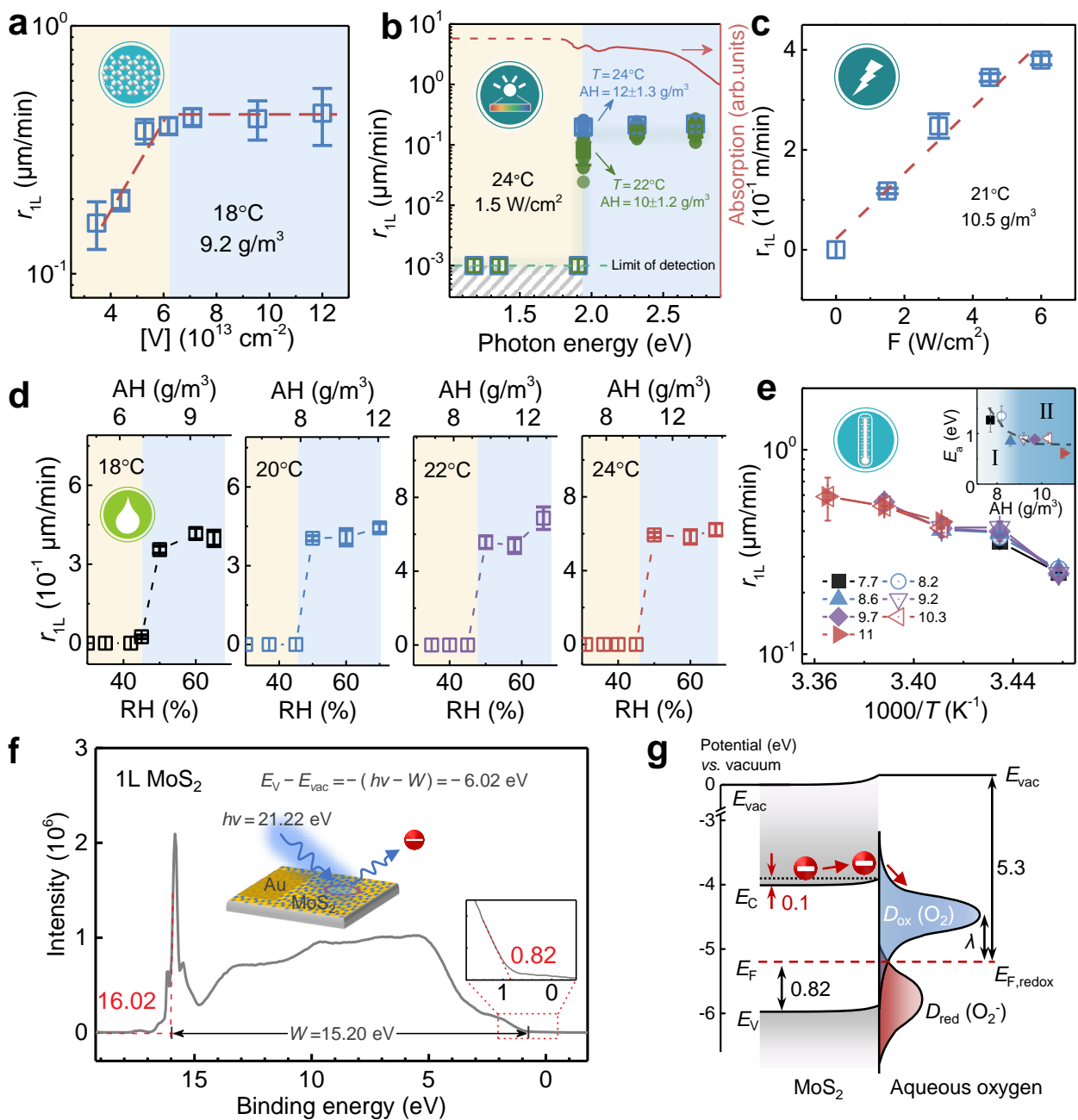

Figure 4